\documentclass[3p, review]{elsarticle}

\biboptions{sort&compress}

\usepackage{bm}
\usepackage{amsmath}
\usepackage{color}

\title{Unraveling the Atomic-Scale Pathways Driving Pressure-Induced Phase Transitions in Silicon}

\author[1]{Fabrizio Rovaris\corref{cor1}}
\ead{fabrizio.rovaris@unimib.it}

\author[1]{Anna Marzegalli}
\author[1]{Francesco Montalenti}
\author[1]{Emilio Scalise}

\affiliation[1]{organization={Department of Materials Science. University of Milano-Bicocca},
    addressline={Via R. Cozzi 55},
    postcode={I-20125},
    city={Milano},
    country={Italy}
}

\date{}

\begin{document}


\begin{abstract}
Silicon exhibits several metastable allotropes which recently attracted attention in the quest for materials with superior (e.g. optical) properties, compatible with Si technology.
In this work we shed light on the atomic-scale mechanisms leading to phase transformations in Si under pressure. To do so, we synergically exploit different state-of-the-art approaches. In particular, we use the advanced GAP interatomic potential both in NPT molecular dynamics simulations and in solid-state nudged elastic band calculations, validating  our predictions with ab initio DFT calculations.

We provide a link between evidence reported in experimental nanoindentation literature and simulation results. Particular attention is dedicated to the investigation of atomistic transition paths allowing for the transformation between BC8/R8 phases to the \emph{hd} one under pure annealing. In this case we show a direct simulation of the local nucleation of the hexagonal phase in a BC8/R8 matrix and its corresponding atomic-scale mechanism extracted by the use of SS-NEB. We extend our study investigating the effect of pressure on the nucleation barrier, providing an argument for explaining the heterogeneous nucleation of the \emph{hd} phase observed in experiments reported in the literature.
\end{abstract}

\maketitle

\section{Introduction}\label{sec:intro}

Silicon is the most important material for the semiconductor industry and its leading role in microelectronics has not been challenged in decades. While silicon has been exploited so far mainly in its most stable crystalline phase, the cubic diamond (\emph{dc}), it also exhibits a wide range of allotropes that can be synthesized by pressure-induced Phase Transitions (PTs). Some of these allotropes, such as the BC8/R8 phases~\cite{MalonePRB2008ii,ZhangPRL2017,WongPRL2019}, the hexagonal diamond (\emph{hd})~\cite{HaugeNanolett2015,PandolfiNanoLetters2018}, and also the amorphous~\cite{BarthChemMat2020,MalonePRB2008} show very interesting optoelectronic properties and can thus be considered as possible candidates to directly integrate different functionalities into Si microelectronics. 

Several Si allotropes have been synthesized inducing solid-solid phase transitions by applying external loadings, typically achieved in experiments by the use of nanoindentation~\cite{BradbyJMR2001,KailerJAP1997,JangActaMat2005}, Diamond Anvil Cells~\cite{HustonNanoLett2021} or laser-driven shock compression~\cite{PandolfiNatComm2022}. We will mainly focus our study on the PTs induced by nano-indentation for its wide diffusion and the completeness of available experimental data~\cite{KiranBOOK2015}, but the simulation methods exploited here may be easily adapted to different loading conditions. The general picture regarding Si PTs in nanoindentation is the following: The cubic diamond structure undergoes a first PT under compressive loading that ends in a metallic phase with the $\beta$-Sn structure that is about 22\% denser than \emph{dc}~\cite{KiranBOOK2015,GerbigJMR2015}. During unloading, silicon does not transform back to \emph{dc} phase. Instead, depending on the specific unloading conditions, different metastable phases such as BC8 or R8, or amorphous silicon are formed~\cite{BradbyJMR2001,DomnichRevAdvMatSci2002}. Finally, a peculiar behavior is observed with annealing of the BC8/R8 phase, with the system not fully recrystallizing into the most stable~\emph{dc} phase but undergoing a transition to the hexagonal diamond phase when annealed at a high enough temperature (starting from about 200~$^{\circ}$C)~\cite{LiangScriptaMat2022,PandolfiNanoLetters2018}. The \emph{hd} phase is metastable and it reverts to \emph{dc} only after annealing at very high temperature, exceeding 700~$^{\circ}$C~\cite{WongJAP2019}.

Classical Molecular Dynamics (MD) simulations may be used to capture the PTs processes but are limited by the space/time scale achievable in simulations and by the accuracy of the potential used to describe atomic interactions. Recent advances in Machine Learning overcome the latter limitation: given the technological and scientific relevance of silicon, some of the most comprehensive interatomic potentials developed so far have been specifically addressed for silicon, like the Gaussian Approximation Potentials (GAP) of Bartok~\emph{et al.}~\cite{bartok2018machine} and the performant atomic cluster expansion (PACE) of Lysogorskiy~\emph{et al.}~\cite{LysogorskiyNPJCompMat2021}. However, the limitations in time/space scale affecting MD simulations still prevent the realistic modeling of PTs. Although we recently demonstrated that direct, large-scale, ML-based simulations of the indentation process may provide hints on the PTs and overcome some limitations of older approaches~\cite{GeActaMat2024}, MD is still only capable of reaching space/time scales that are 3/4 orders of magnitude smaller and 6/7 orders of magnitude faster than a real experiment~\cite{KiranBOOK2015}.

A more versatile simulation technique commonly used to model PTs is the Solid-State Nudged Elastic Band (SS-NEB)~\cite{SheppardJCP2012}. This technique generalizes the NEB method~\cite{HenkelmanJCP2000} to the description of solid-state phase transitions, allowing the study of Minimum Energy Paths (MEPs) connecting different meta-stable phases. In this approach, small periodic cells of the phases under consideration are typically used to capture the transition path. This approach is computationally very convenient due to the limited number of atoms involved in the calculations (typically of the order of the unit cell or small replica of that) and provides an easy way to estimate the kinetic barriers. The small size of the simulation cells, however, shows some limitations when attempting to compare with experimental results. Indeed, the choice of the initial and final simulation cells is not unique and thus the MEP found cannot be considered as the global minimum energy path. Moreover, the energy barrier describing the transition of a small cell cannot be scaled to the size of the experiments, because it would involve the simultaneous transition of the whole system towards the new phase due to periodicity. SS-NEB  calculations could in principle be performed on larger systems but they become increasingly more computationally demanding and the initial guess for the transition path becomes impracticable without any \emph{a priori} knowledge of the process. 

Nucleation and growth of the new crystal phase is expected to be the main mechanism of phase transition in real experimental conditions, but it is still challenging to address it quantitatively and predictively in simulations. Unraveling the atomic-scale process underlying pressure-induced PTs is of paramount importance to understand the parameters (stress, temperature, boundary conditions,\dots) influencing the microstructure formation of meta-stable phases. In this paper we show how a combined use of SS-NEB and MD-NPT can be exploited to overcome the main limitations of the two approaches, tackling the problem of modeling the nucleation and growth of a new phase in a solid-solid transition process. We address the atomic-scale processes at the bases of all the PTs encountered during Si nanoindentation as an example of the synergic use of these two techniques, but the simulation approach presented here may be easily extended to different experimental conditions. The paper is organized as follows: in Section~\ref{sec:methods} we introduce the computational methods exploited in this work. In Section~\ref{sec:path} we propose a full transition path, found by SS-NEB, starting from the \emph{dc} phase and linking all the metastable phases encountered during an indentation process. In the subsequent section, we detail each transition process encountered, complementing the SS-NEB results with the use of pressure-controlled MD simulations. We investigate in detail the effect of applied external stress fields and we check the validity of an analytical formulation for describing the pressure dependence of the energy barriers. Finally, particular focus is devoted to the study of a local nucleation process, detailed in Section~\ref{sec:local}: the transition to the \emph{hd} phase under annealing of the BC8/R8 phases. 

\section{Computational Methods}\label{sec:methods}

MD simulations have been performed with the LAMMPS code~\cite{ThompsonCPC2022}. The interatomic potentials used for silicon is the ML-based GAP potential~\cite{bartok2018machine}. The atomic configurations are visualized and analyzed using the OVITO software~\cite{stukowski2009visualization} and a Neural Network-based tool that we proposed in a recent work~\cite{GeActaMat2024} was used for phase recognition. Pressure-controlled simulations were performed in the NPT ensemble by applying a Nosé-Hoover thermostat and a barostat based on the Parrinello-Rahman approach~\cite{ParrinelloJAP1981}. The timestep considered in all the simulations was $1~\text{fs}$ and each NPT-MD run was conducted by allowing the system to relax and thermalize at zero external stress for $10000$ integration steps before applying any external loading. 

SS-NEB calculations have been performed with the TSASE simulation framework~\cite{TSASE}, considering the climbing image method~\cite{HenkelmanJCP2000_2} in order to better approximate the value of the kinetic barriers. The SS-NEB calculations were considered converged when the maximum force on the images along the path reached a value equal or smaller than $10^{-3}~\text{eV/\AA}^3$ when considering the small periodic cells. The MEP reported in Section~\ref{sec:local} for the 432 atom cell showing the nucleation process has been converged up to $10^{-2}~\text{eV/\AA}^3$. The search for atomic correspondence in some of the MEPs found for the small periodic cells was performed by exploiting a search tool based on the minimization of the geometric distance traveled by the atoms along the transition path, as described in Ref.~\cite{TherrienPRL2020}. Specifically, the $\beta$-Sn to BC8 transition was found exploiting this tool.

\section{Full transition path in Si nanoindentation}
\label{sec:path}

\begin{figure}[tbh]
    \centering
	\includegraphics[width=0.80\textwidth]{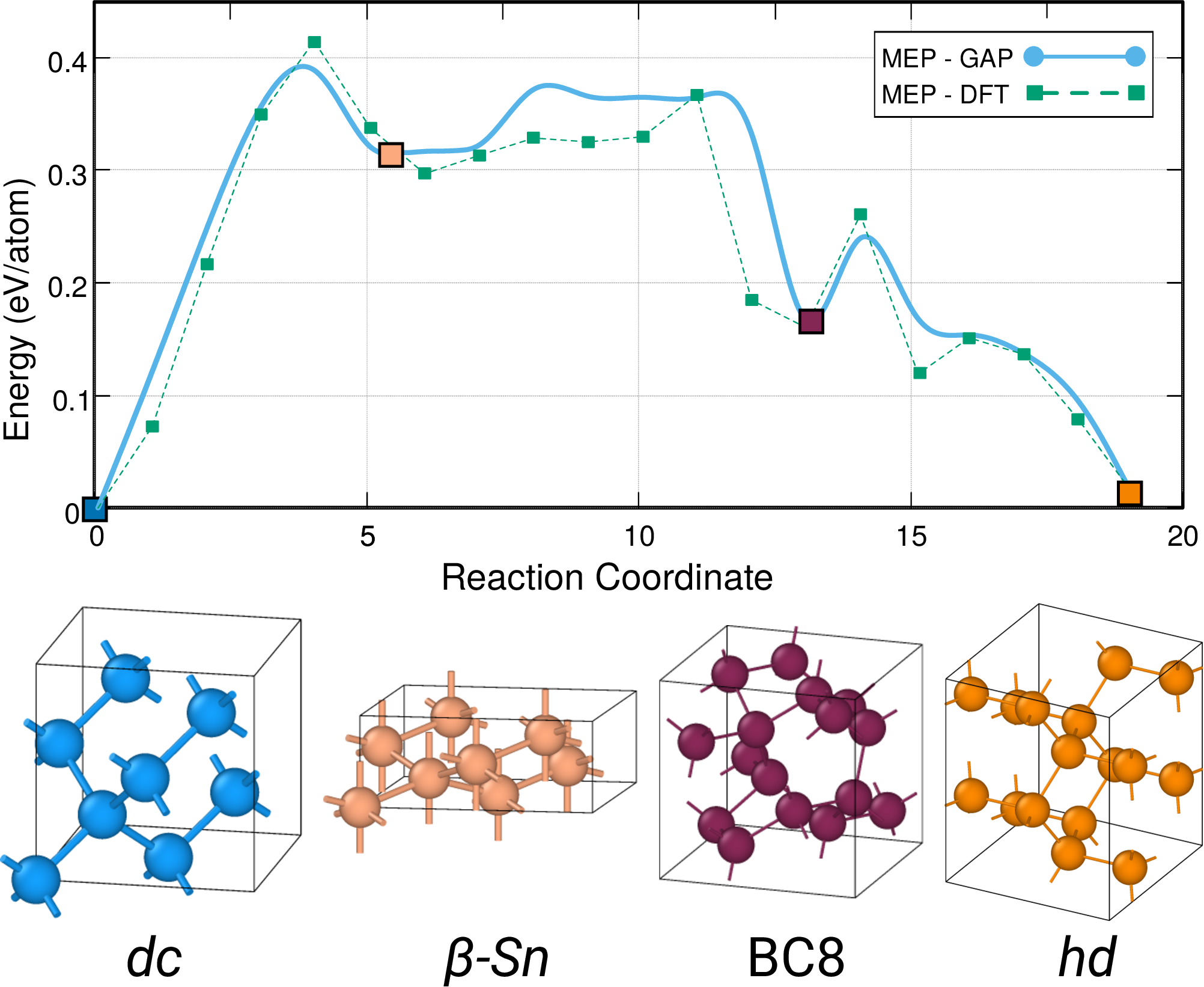}
	\caption{Complete transition path connecting all the metastable phases encountered during Si nanoindentation and successive annealing, as calculated both by GAP and DFT. From left to right: diamond cubic (\emph{dc}), $\beta$-Sn, BC8 and hexagonal diamond (\emph{hd}).}
	\label{fig:path}
\end{figure}

Phase Transitions can be conveniently modeled by SS-NEB as mentioned in the Introduction. In this approach, small simulation cells of the phases under consideration are connected by MEPs parametrized by a Reaction Coordinate that draws the transition path in the high dimensional Potential Energy Surface. However, as mentioned in the Introduction, the choice of the (periodic) supercells used as initial and end points in the SS-NEB calculation is not unique and thus one can never be sure that the global MEP has been found. Following transition state theory, the rate $\Gamma$ of a specific transition is connected to its kinetic barrier $\Delta E$ by: 
\begin{equation}
    \Gamma = \tilde{\nu} e^{-\dfrac{\Delta E}{k_{\text{b}}T}}
\label{eq:rate}
\end{equation}
where, $\tilde{\nu}$ is a frequency prefactor, $k_{\text{b}}$ is the Boltzmann constant and $T$ the absolute temperature~\cite{VineyardJPCS1957}. The transition rate is exponentially dependent on the kinetic barrier and thus the MEP exhibiting the smallest barrier value can be considered the most accurate atomic-scale description of the transition process. In Fig.~\ref{fig:path} we illustrate an application of SS-NEB by selecting nanoindentation (followed by annealing) as a relevant physical process for inducing phase transitions, as briefly described in the Introduction. We thus connected all the phases encountered during a typical nanoindentation experiment and we report a unique transition path found by SS-NEB. Results for both the GAP potential and density fuctional theory (DFT) calculations are reported. We would like to stress that the DFT results reported here are fully-converged MEPs and not single point calculations repeated on the transition images found by the use of the GAP potential. The good overall agreement obtained by the use of GAP and DFT strongly assesses the predictive ability of the GAP potential, confirming the accuracy of its kinetic predictions in addition to the already widely investigated equilibrium properties~\cite{bartok2018machine}. In the Supplementary Material of this paper we further expand this validation by a direct comparison of the Potential Energy Surface (PES) neighborhood of some of the phases here considered (Fig. S1) and with additional MEPs comparisons.

The path reported in Fig.~\ref{fig:path} is composed of three individual transition paths: a first \emph{dc} to $\beta$-Sn transition with an energy barrier of about $396~\text{meV/atom}$, a second transition from $\beta$-Sn to the BC8 phase with an energy barrier of about  $68.2~\text{meV/atom}$ and a final transition to the \emph{hd} phase with an energy barrier of about $72.5~\text{meV/atom}$. The BC8 phase is often found in coexistence with another similar phase, named R8. We thus investigated also the transition path between these two phases and we found a small kinetic barrier amounting to about $29.6$~ meV/atom. This small kinetic barrier is of the order of the thermal energy of the atoms at room temperature and the transition path found involves only one main atomic movement per unit cell. For these reasons, as we will better elucidate in the following of the paper, we can consider the BC8 and R8 phases as co-existing at the typical experimental conditions. This behavior will be discussed in more details in Section~\ref{sec:load_unload}. 

All the MEPs composing the complete path have been separate objectives of different research works in the past years. In fact, alternative paths were reported in the literature for the transition $\beta$-Sn to BC8 in Ref.~\cite{YinnpjCompMat2020} and the transition BC8 to \emph{hd} in Refs.~\cite{WangPRL2013}. Still, the path reported in Fig.~\ref{fig:path} can be considered to be the best representative transition path connecting all the phases considered here, because the kinetic barriers found in this work are lower than any analogous value reported in the literature, to the best of our knowledge. However, as already mentioned several times, the atomic-scale mechanisms underlying Si nanoindentation are still not clear. Arbitrariness in the choice of the simulation cells prevents the generalization of the results of Fig.~\ref{fig:path}. For this reason, we will focus separately on each specific transition, investigating the atomic-scale mechanism with the combined use of SS-NEB and NPT-MD simulations. 

\section{Phase transitions during loading and unloading}
\label{sec:load_unload}

\begin{figure}[tbh]
    \centering
	\includegraphics[width=1.0\textwidth]{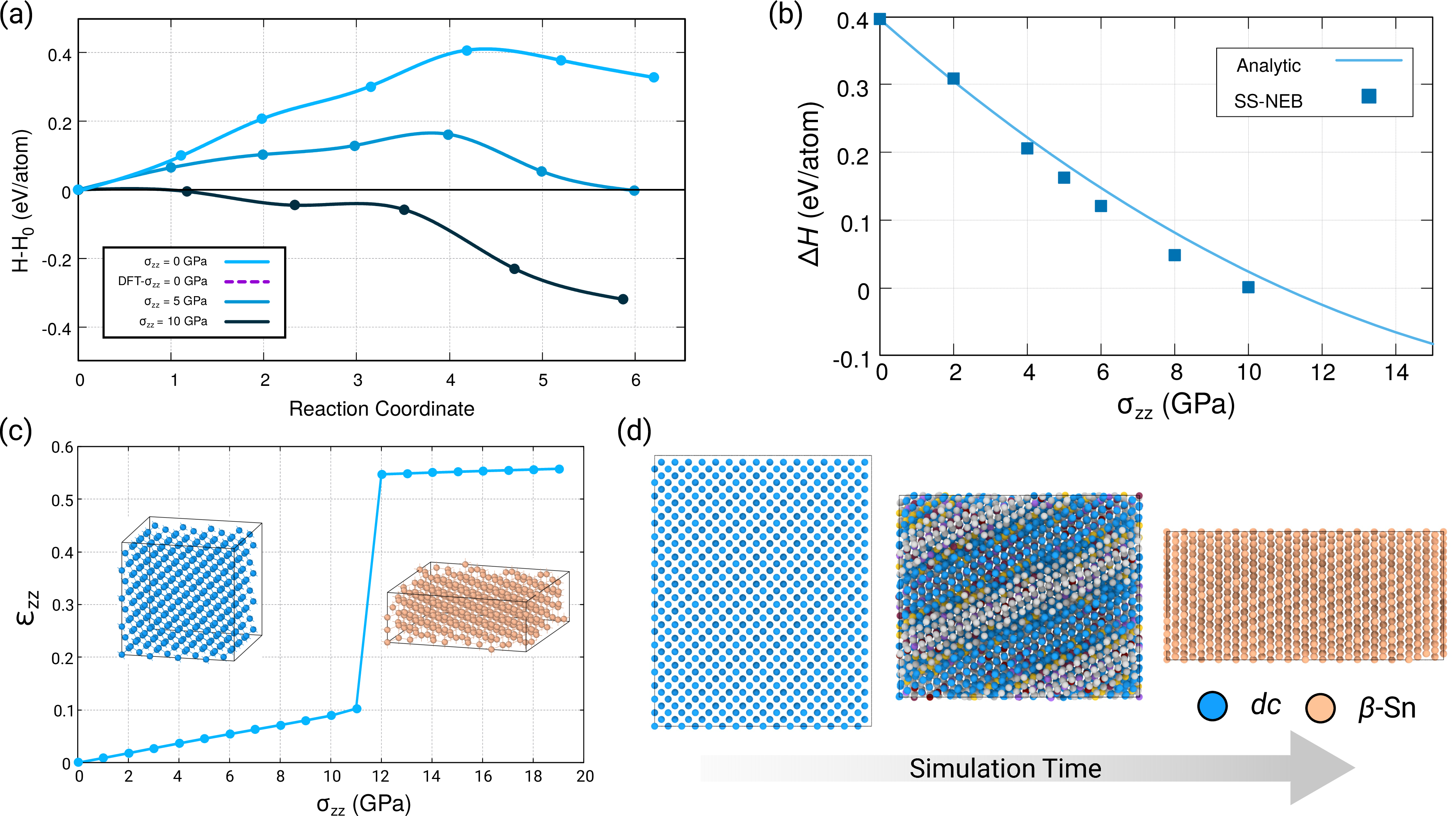}
	\caption{(a) MEPs calculated under an applied uniaxial pressure $\sigma_{zz}$ for the \emph{dc} to $\beta$-Sn transition. (b) Barrier value vs applied uniaxial stress as calculated by SS-NEB method and the analytical expression. (c) Strain vs Stress plot showing a sharp discontinuity in the $\varepsilon_{zz}$ value at the transition. (d) Snapshots of an NPT-MD simulation of the \emph{dc} to $\beta$-Sn transition showing band-like structures due to the reorganization of the internal stress.}
	\label{fig:betasn}
\end{figure}

Silicon undergoes a phase transition from the \emph{dc} semiconductor phase to the much denser, metallic, $\beta$-Sn phase under compressive stress. This occurs at a contact pressure close to $5.2~\text{GPa}$ in a nanoindentation experiment and can be monitored \emph{in situ} by Raman analysis~\cite{GerbigJMR2015}. The transition has also been reproduced in MD simulations of nanoindentation exploiting both semi-empirical~\cite{JiapengSciRep2017,abram2022silicon} and ML potentials~\cite{GeActaMat2024}. This process can  be investigated by SS-NEB calculations as we have already shown in Fig.~\ref{fig:path}. We further investigated this transition process by repeating SS-NEB calculations with the addition of a compressive uniaxial stress field and we report the corresponding enthalpy barrier $H$ in Fig.~\ref{fig:betasn}(a). In the Figure, we report MEPs found by repeating SS-NEBs calculations with the addition of different values of the applied stress ($\sigma_{zz}$ component, with $z$ being parallel to a $[100]$ direction). We report three different MEPs: $\sigma_{zz} = 0~\text{GPa}$ , $\sigma_{zz} = 5~\text{GPa}$ and $\sigma_{zz} = 10~\text{GPa}$. The MEP calculated without external pressure is equivalent to the one reported in Fig.~\ref{fig:path} (the first part connecting \emph{dc} and $\beta$-Sn); the MEP at $\sigma_{zz} = 5~\text{GPa}$ represents a stress condition where $H_{\text{\emph{dc}}} \approx H_{\text{$\beta$-Sn}}$. The value of $5$~GPa for the thermodynamic stability of the $\beta$-Sn phase is in very close quantitative agreement to the experimental value of $5.2$~GPa obtained by \emph{in situ} Raman analysis performed during nanoindentation~\cite{GerbigJMR2015}. Finally, the path at $\sigma_{zz} = 10~\text{GPa}$ shows a condition when the kinetic barrier for the transition is approximately zero. 

The effect of the external pressure on the kinetic barrier for the transition can also be estimated by exploiting a modified Bell theory~\cite{BellScience1978,GhasemiJMPS2020}. The $\Delta E$ contribution in Eq.~\eqref{eq:rate} used for calculating the transition rate with no applied external stress is substituted by an enthalpy term $\Delta H (\bm{\sigma}^{\text{ext}})$ dependent on the applied external stress $\bm{\sigma}^{\text{ext}}$ by:
\begin{equation}
    \Delta H (\bm{\sigma}^{\text{ext}}) = E^{(t)}(\bm{\sigma}^{\text{ext}})-E^{(0)}(\bm{\sigma}^{\text{ext}}) - V_0 \bm{\sigma}^{\text{ext}}:(\bm{\varepsilon}^{(t)}(\bm{\sigma}^{\text{ext}})-\bm{\varepsilon}^{(o)}(\bm{\sigma}^{\text{ext}}))
    \label{eq:enthalpy}
\end{equation}
where $E^{(o),(t)}(\bm{\sigma}^{\text{ext}})$ are the potential energies of the original (o) and the transition (t) states at the applied external stress, $V_{0}$ is the volume of the original state and $\bm{\varepsilon}^{(o),(t)}(\sigma^{\text{ext}})$ are the Green-Lagrange strain tensors describing the deformation of the original (o) and transition (t) states at the applied external stress. Eq.~\eqref{eq:enthalpy} can be expanded in Taylor series around the zero-stress state $(\bm{\sigma}^{\text{ext}}=\bm{0})$ as in Ref.~\cite{GhasemiJMPS2020} for small deformations, in order to obtain a formulation that depends only on quantities calculated at zero external stress:
\begin{equation}
    \Delta H (\bm{\sigma}^{\text{ext}}) = \Delta E(\bm{0}) - V_0 \bm{\sigma}^{\text{ext}}:(\bm{\varepsilon}^{(t)}(\bm{0})-\bm{\varepsilon}^{(o)}(\bm{0})) - \dfrac{V_0}{2}\bm{\sigma}^{\text{ext}}:\bigg{(}\dfrac{\partial \bm{\varepsilon}^{(t)}(\bm{0})}{\partial \bm{\sigma}^{\text{ext}}}-\dfrac{\partial \bm{\varepsilon}^{(o)}(\bm{0})}{\partial \bm{\sigma}^{\text{ext}}}\bigg{)}:\bm{\sigma}^{\text{ext}}+\mathcal{O}(\bm{\sigma}^2)
    \label{eq:taylor}
\end{equation}
which can be easily read as composed of: the energy barrier at zero stress, a linear term representing the work done by the external stress for deforming the system from the original state to the transition state and a second order term accounting for the difference in elastic response between the original and transition states. 

The pressure dependence for the \emph{dc} to $\beta$-Sn transition is highlighted in Fig.~\ref{fig:betasn}(b) where we plot the enthalpy barrier $\Delta H$ as a function of the applied uniaxial stress $\sigma_{zz}$. We report both the results obtained by repeating SS-NEB calculations and by exploiting Eq.~\eqref{eq:taylor} with a second order approximation of the Taylor expansion. The analytical prediction is found to be in good agreement, providing a convenient approach to predict the pressure dependence of energy barriers.

The first order nature of the phase transition between the \emph{dc} and $\beta$-Sn phases is highlighted in Fig.~\ref{fig:betasn}(c) where we report the strain vs stress curve obtained by relaxing the \emph{dc} structure at each given uniaxial stress value. The curve shows a sharp discontinuity in the $\varepsilon_{zz}$ component at the transition pressure. The same process has also been investigated with MD-NPT simulations, in which an applied field $\sigma_{zz}$ has been increased during the simulation from zero to $12~\text{GPa}$. The transition happens, consistently with Refs.~\cite{LevitasPRB2017,LevitasPRL2017}, as soon as a lattice instability condition is reached. At this condition, the \emph{dc} simulation cell transforms into the $\beta$-Sn phase heterogeneously due to the presence of perturbations caused by thermal fluctuations. This heterogeneity is evident in the intermediate snapshot of the simulation shown in Fig.~\ref{fig:betasn}(d), where residual \emph{dc} is structured into band-like shapes due to the reorganization of the internal stress. In the final snapshot shown, a stable state of the $\beta$-Sn phase is formed. 

\begin{figure}[tbh]
    \centering
	\includegraphics[width=1.0\textwidth]{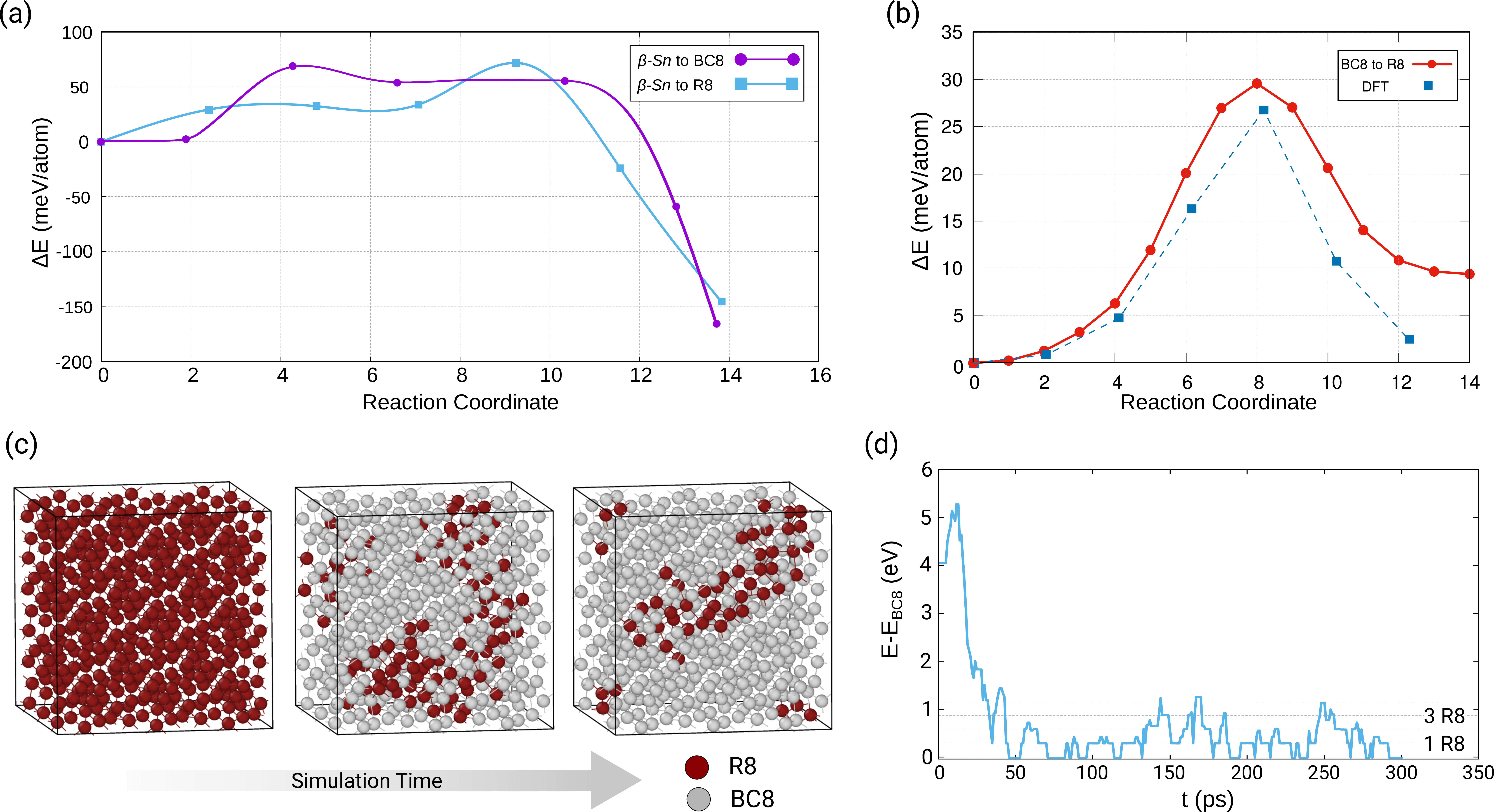}
	\caption{(a) Minimum Energy Paths connecting the $\beta$-Sn phase to the metastable phases BC8 and R8. (b)  MEPs for the BC8 to R8 transition and the same path converged by DFT. (c) MD-NPT simulation of the coexistence of BC8 and R8 phases. The simulation starts from a 432-atom cell of the R8 phase  and shows the transition to a BC8/R8 mixture. (d) Energy (relative to BC8) of relaxed configurations taken during the MD simulation of panel (c).}
	\label{fig:BC8_R8}
\end{figure}

The $\beta$-Sn phase is stable under applied external pressure but typically shows a transition to the metastable BC8 and R8 phases when the external load is released. During a nanoindentation experiment this PT happens during the unloading stage and is evidenced in several studies~\cite{SchaffarJOM2022}. The formation of the two metastable phases depends on the unloading speed and is evidenced during slow enough unloading conditions~\cite{LinPRL2020}.  In all the conditions a mixture of BC8 and R8 phases is always present after full unloading, while the ratio between the two phases cannot be controlled and is dependent on the processing temperature~\cite{MannepalliJAP2019}. In Fig.~\ref{fig:BC8_R8}(a) we explain this phenomenon. As can be seen in the figure, we calculated the kinetic barrier by SS-NEB both for the transition from $\beta$-Sn to BC8 and to R8 (the images along the paths found for both the transitions are reported in the Supplementary Material, figure S3 and S4, respectively).. The two energetic barriers are indeed very similar and amount to $68.2~\text{meV/atom}$ and $72.5~\text{meV/atom}$, respectively. The $\approx 4~\text{meV/atom}$ energy difference is so small that the two transition processes have a very similar transition rates and thus a mixture of the two phases is always encountered after the full unloading of the indentation tip. Another explanation for the simultaneous presence of the two phases is the result reported in Fig.~\ref{fig:BC8_R8}(b). The MEP reported in the figure for the transition between these two phases shows an energy barrier of about $29.6$~ meV/atom. In the figure we also compare the MEP found with a SS-NEB calculation performed by first principle DFT. Remarkably, the same MEP can be converged also with DFT and shows a very similar kinetic barrier of $26.7$~ meV/atom, confirming the accuracy of the entire kinetic of the transition process. Moreover in the transition path found (fully reported in the Supplementary Material, figure S5), only one main atomic movement per unit cell is involved. For these reasons these two phases should always appear as a mixture at equilibrium, with their relative ratio being only dependent on the temperature. This finding is consistent with the observations of Ref.~\cite{PiltzPRB1995} where the authors estimated an upper limit to the kinetic barrier of about $40~\text{meV/atom}$. 

Moreover, we complemented our SS-NEB calculations by simulating the co-existence between these two phases in a MD-NPT simulation. In Fig.~\ref{fig:BC8_R8}(c) we show a simulation starting from a pure R8 phase (whose atoms are colored in red) containing $432$ atoms and annealed at $700~\text{K}$. In the snapshots shown (a full movie of the simulation can be found in Supplementary Movie 1) the R8 phase quickly transforms into a mixture containing mainly the BC8 phase (which is found to be slightly more energetically favoured according to GAP calculations) and proceeds with random phase transitions between BC8 and R8 inside the simulation cell. This behavior can be better appreciated in the plot reported in Fig.~\ref{fig:BC8_R8}(d). In the panel we report the energy values (relative to the energy of the BC8 phase) of minimized snapshots taken along the simulation reported in Fig.~\ref{fig:BC8_R8}(c). As it can be observed in the plot, the energy of the simulation cell quickly converges towards the energy of a full BC8 simulation cell (zero value in the plot) with multiple discrete jumps visible during the simulation time that can be easily correlated with the formation of R8 units cells inside a matrix mainly consisting of BC8.  

Based on our results we can argue that a mixture of the BC8 and R8 phases is always naturally formed after the unloading of the $\beta$-Sn phase for a couple of reasons: first, a very similar transition barrier has been found for the transition between the $\beta$-Sn phase and both BC8 and R8 (see Fig.~\ref{fig:BC8_R8}(a)); second, these two latter phases show very similar formation energy (even closer in DFT than GAP potential, as evident in Fig~\ref{fig:BC8_R8}(b)) and have very closely related crystalline structure. This implies that a very simple transition path exists between them (involving only one main atomic movement) with a kinetic barrier compatible with the thermal energy (see Fig.~\ref{fig:BC8_R8}(b)). For these two reasons we expect the two phases to co-exist at typical experimental conditions, similarly with the MD simulation shown in Fig.~\ref{fig:BC8_R8}(c), and consistently with experimental observations done after the unloading step in nanoindentation experiments~\cite{SchaffarJOM2022}. 
 
\section{Local Nucleation for the BC8/R8 to \emph{hd} transition}
\label{sec:local}

\begin{figure}[tbh]
    \centering
	\includegraphics[width=1.0\textwidth]{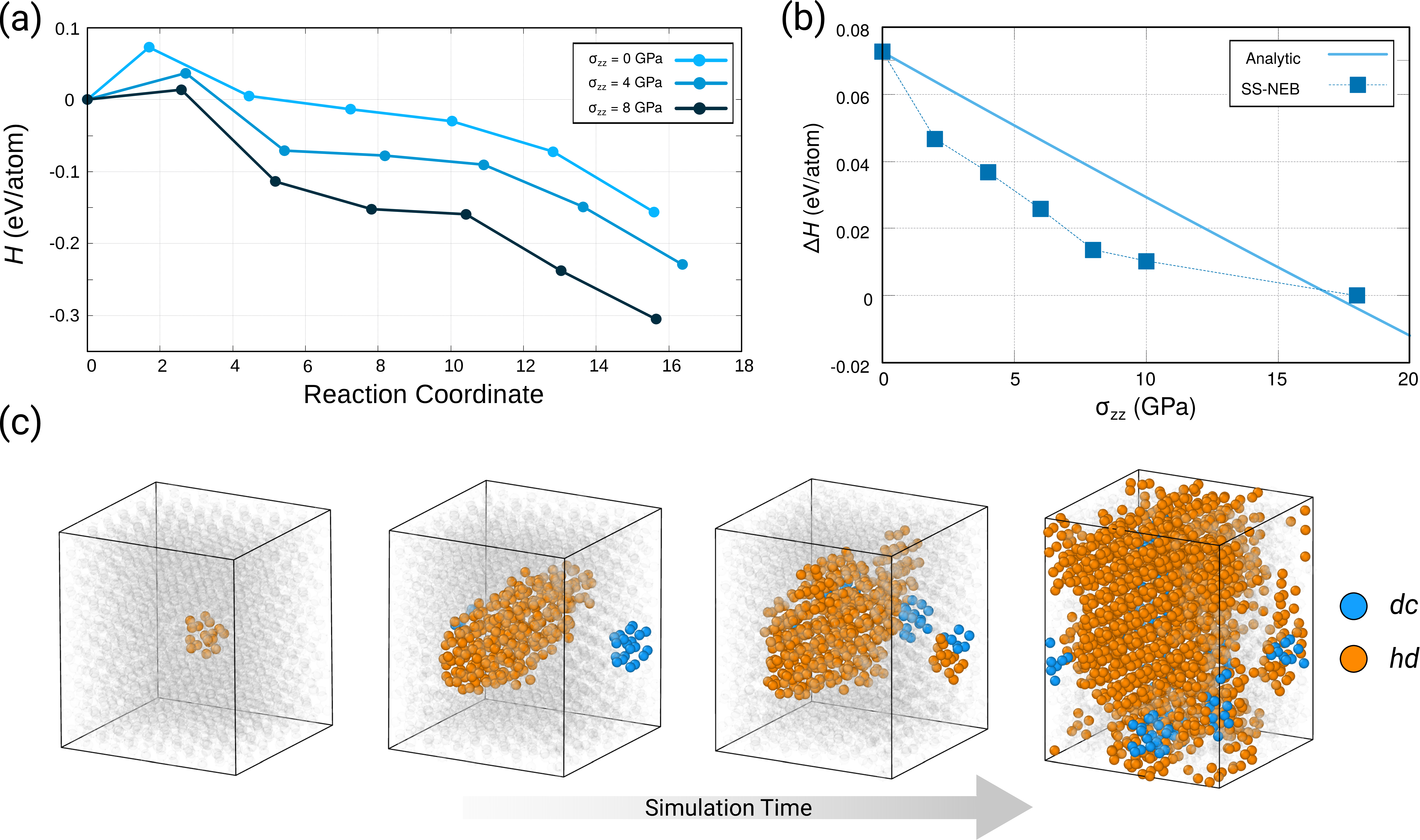}
	\caption{(a) MEPs for the BC8 to \emph{hd} transition calculated with no applied pressure and with uniaxial tensile stress. (b) Value of the kinetic barrier as a function of the applied uniaxial stress ($zz$ component) for the BC8 to HD transition, comparison between SS-NEB  and analytical results. (c)  Nucleation mechanisms of the hexagonal diamond phase in a BC8 matrix as observed in one of the NPT simulations of annealing at $T=700~\text{K}$.}
	\label{fig:NPT}
\end{figure}

Finally, we focus on the transition towards the \emph{hd} phase after the annealing of the BC8/R8 phases. In this case, the energy barrier for the transition, reported in Fig.~\ref{fig:path}, amounts to $\approx 72~\text{meV/atom}$. The hexagonal diamond phase is less dense than the BC8 phase and its conventional cell is expanded in the vertical direction with respect to the BC8 cell. This suggests a possible transition path involving an expansion in the vertical direction during the annealing (the images along the transition path found are reported in the Supplementary Material, figure S6). This hypothesis has been confirmed by repeating the SS-NEB calculation with the application of uniaxial tensile stress in the vertical direction. As shown in Fig.~\ref{fig:NPT}(a) we investigated different values of this uniaxial pressure and observed a strong decrease in the kinetic barrier for the transition. In Fig.~\ref{fig:BC8_R8}(b) we report the value of the kinetic barrier at different tensile uniaxial stress as calculated by SS-NEB and by the analytical expression of Eq.~\eqref{eq:taylor}. The analytical prediction is not reproducing so closely the simulated values in this case. However, a strong dependence of the barrier on the uniaxial stress is observed in both cases, with its value vanishing around $18~\text{GPa}$, as extracted by SS-NEB calculations. 

The above observation suggests a possible a possible pathway for speeding up the direct observation of the transformation using MD simulations. We thus initialized a large simulation cell of the BC8 phase consisting of $3456$ atoms. We mimicked a progressive uniaxial tensile stress on the simulation cell by imposing a strain rate $\dot{\varepsilon}_{zz}$ and barostating to zero all the stress components except the $\sigma_{zz}$ during the NPT simulation. This procedure, albeit artificially used here for accelerating the time scale of the transition process, may be justified on a physical basis by analyzing the results of indentation experiments. The matrix of metastable BC8/R8 phases is indeed experiencing a strain effect from the un-transformed \emph{dc} substrate around it. This effect has been experimentally verified by annealing metastable phases obtained by nanoindentation in a \emph{dc} matrix or in an amorphous one. In Refs.~\cite{RuffellAPL2007,RuffellJAP2009} the authors found that the presence of a \emph{dc} matrix surrounding the metastable phases accelerates the transition process. This observation does provide a physical motivation for the imposition of uniaxial stress in the vertical direction in our simulation in order to destabilize the BC8/R8 phases out of their ground state. We thus simulated the annealing of a BC8 matrix and tracked the evolution of the stress and phases during the simulation. Selected snapshots of the simulation (the full simulation movie is reported in the Supplementary Material as Supplementary Movie 2) are reported in Fig.~\ref{fig:NPT}(c) for the annealing temperature of $T = 700~\text{K}$. In the figure, the BC8 atoms are colored in gray, while the \emph{hd} and \emph{dc} are in orange and blue, respectively. As it can be observed a small nucleus of the \emph{hd} phase is formed first, followed by a growth at a later stage of the simulation. A second seed of \emph{dc} can also be seen in the simulations, directly showing the competition between the formation of \emph{hd} or \emph{dc} in annealing experiments. Thus, despite we initialized the simulation cell and the loading condition based on results taken from the MEPs under pressure on a small periodic cell, the mechanisms of formation of the \emph{hd} phase in the MD-NPT simulation can be followed as a nucleation and growth of a \emph{hd} seed in the BC8 matrix.

\begin{figure}[tbh]
    \centering
	\includegraphics[width=1.0\textwidth]{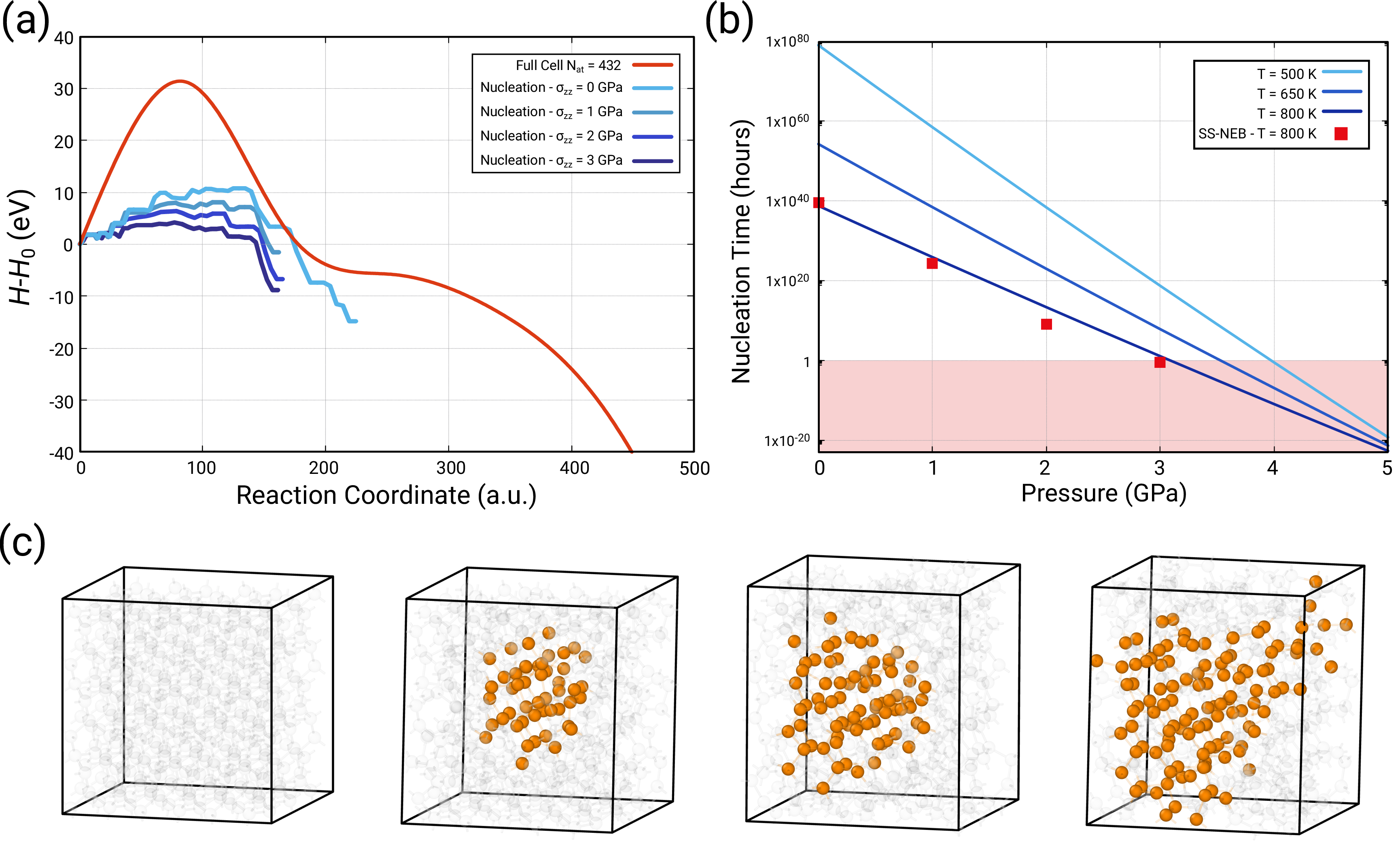}
	\caption{(a) Minimum Energy Path found for the nucleation mechanisms of the hexagonal phase inside a BC8 matrix in a simulation cell containing 432 atoms as a function of the applied tensile uniaxial pressure $\sigma_{zz}$. The total energy barrier for the full transition of a simulation cell containing the same number of atoms is shown as a reference in red. (b) Plot of the expected nucleation time (logarithmic scale) as a function of the pressure for a realistic-sized system. (c) Images taken along the minimum energy path of the nucleation process. }
	\label{fig:local}
\end{figure}

More interestingly, based on the results of the MD-NPT simulation, one has the possibility to model the atomic-scale process of the transition between the BC8 and \emph{hd} phase via local nucleation. Then, using SS-NEB, a more realistic estimation of the kinetic barrier can be obtained, avoiding all the limitations of MEPs found on small periodic cells, as discussed in the Introduction. Moreover, the kinetic barrier for the local nucleation process computed by SS-NEB can now be calculated at any applied external pressure, providing a possible way to predict the stress condition for which the transition can be experimentally observed. First, we selected a smaller region from the simulation of Fig.~\ref{fig:NPT}(c) containing 432 atoms in the cell and we computed the energy barrier for the BC8 to \emph{hd} phase via local nucleation. In order to perform SS-NEB calculation on this cell we split the calculation into 8 different SS-NEB paths with 7 images each. The convergence criterion is met when the maximum force on the images is below $10^{-2}~\text{eV/\AA}$. After this, we repeated the SS-NEB calculation with the addition of several values of external uniaxial tensile pressures. The final results are shown in Fig.~\ref{fig:local}(a) where the nucleation barrier calculated in this way is compared to the barrier for the full periodic cell transition with an equivalent number of atoms. As can be seen in the figure, the barrier for local nucleation is several eV smaller than the full cell transition with this specific choice of simulation cell. Moreover, the full cell transition will scale linearly with the number of atoms in the simulation cell, making the total energy value of the kinetic barrier for the process quickly unphysical. On the contrary, the barrier for the local nucleation event is not dependent on the cell size and thus can be considered a more realistic estimation of the atomic-scale transition process. 

Moreover, as can be seen in figure~\ref{fig:local}(a), we found a strong dependence of the barrier on the external stress. The value of the kinetic barrier for the nucleation is lowered from about $10.7~\text{eV}$ at zero external pressure to about $4.4~\text{eV}$ at $3~\text{GPa}$ of tensile stress. This result suggests a possible way to interpret experimental results where the \emph{hd} phase obtained after annealing of the BC8/R8 phases is structured into small crystals sized a few tens of nanometers~\cite{GeJAP2004}. Based on our results we argue that, due to the exponential dependence of the transition rate on the kinetic barrier (see Eq.~\eqref{eq:rate}), a heterogeneous residual stress field in the indentation region may play a role in the formation of the nanostructured \emph{hd} phase.

In order to better illustrate this point we reported in Fig.~\ref{fig:local}(b) the average transition time as a function of the tensile uniaxial stress $\sigma_{zz}$. Despite the calculation of transition times for nucleation events and the direct comparison with experimental results is universally recognized as challenging~\cite{BlowJCP2021} due to uncertainties in both the numerical prediction (frequency prefactors, finite-size effects, accuracy of the interatomic potentials, etc.) and the lack of direct experimental probes for the nucleation, we attempted an indicative comparison. We estimated the total nucleation rate as the product of the individual nucleation rate (Eq.~\eqref{eq:rate}) with the number of possible nucleation sites, here roughly estimated as the number of atoms in the matrix phase. We considered for a quick estimate a typical value for the volume of the transformed region in a nanoindentation experiment ($2~\text{$\mu$m}$ radius for $1~\text{$\mu$m}$ height spherical cap), annealing temperatures between 600 and 800 K and a frequency prefactor of the order of $1\times10^{13}~\text{1/s}$. The line plots in Fig.~\ref{fig:local}(b) have been obtained by predicting the energy barrier at different stress values by using Eq.~\eqref{eq:taylor} and using these values for evaluating  the transition times from Eq.~\eqref{eq:rate}. The red-shaded region corresponds to a typical experimental timescale of $1$ hour annealing time. As can be seen in the figure, a realistic transition time for the nucleation process is obtained at stress values between $3~\text{and}~4~\text{GPa}$, depending on the annealing temperature, giving a possible explanation for the heterogeneous phase transition experimentally observed. The stress values calculated by direct SS-NEB calculations (for $\sigma_{zz} = 0,1,2,3$~GPa, corresponding to the MEPs reported in panel (a)) have also been used to estimate the transition time for $T=800~\text{K}$ and are reported as red squares in the plot, confirming the values estimated by Eq~\eqref{eq:taylor}. While we cannot guarantee that the here-proposed mechanism is exactly the one that leads to the observed transition, and others with lower associated barriers could exist, we find it remarkable to have found a path with an associated rate possibly compatible with the experimental observations.

\section{Conclusion}

In this paper we reported a detailed investigation of pressure-induced PTs in Si exploiting a synergic use of SS-NEB and NPT-MD simulations based on a ML interatomic potential. We showed how these two techniques can be combined in order to overcome the main limitations of the two separate approaches and obtain a more realistic description of pressure-induced phase transitions.

First, we reported a complete transition path linking the \emph{dc} to \emph{hd} phase encompassing all the metastable phases encountered during Si nanoindentation and we detailed each step in the full transition path and the atomic-scale processes there involved. We reproduced the lattice instability involved during the \emph{dc} to $\beta$-Sn transition and we obtained a good agreement for the transition pressure extracted by SS-NEB and experiments. We also confirmed the validity of the modified Bell theory for the prediction of the stress dependence of kinetic barriers. Then, we provided an explanation for the transition during the unloading of the $\beta$-Sn phase to a mixture of the BC8/R8 by a critical comparison of the kinetic barriers and by showing the coexistence of these two phases at the typical experimental temperatures. 

Finally, we notably showed a local nucleation process of the \emph{hd} phase inside a matrix of BC8/R8 under annealing by the combined use of MD-NPT simulations and SS-NEB calculations. We used this result to directly show a solid-solid transition process with the nucleation and growth mechanism. We demonstrated the dependence of the nucleation barrier on applied external stress and provided an argument for the heterogeneous nucleation of the hexagonal phase due to the non-uniformity of residual stresses in a typical nanoindentation experiment. 

\section*{Acknowledgements}
The authors acknowledge the CINECA consortium under the ISCRA initiative for the availability of high-performance computing resources and support.
E.S. acknowledges financial support under the National Recovery and Resilience Plan (NRRP), Mission 4, Component 2, Investment 1.1, Call for tender No. 104 published on 2.2.2022 by the Italian Ministry of University and Research (MUR), funded by the European Union – NextGenerationEU– Project Title "SiGe Hexagonal Diamond Phase by nanoIndenTation (HD- PIT)" – CUP H53D23000780001 - Grant Assignment Decree No. 957 adopted on 30.06.2023 by the Italian Ministry of Ministry of University and Research (MUR).

\bibliographystyle{unsrt}
\bibliography{biblio}

\begin{thebibliography}{10}

\bibitem{MalonePRB2008ii}
Brad~D. Malone, Jay~D. Sau, and Marvin~L. Cohen.
\newblock Ab initio study of the optical properties of {Si-XII}.
\newblock {\em Phys. Rev. B}, 78:161202, Oct 2008.

\bibitem{ZhangPRL2017}
Haidong Zhang, Hanyu Liu, Kaya Wei, Oleksandr~O. Kurakevych, Yann Le~Godec,
  Zhenxian Liu, Joshua Martin, Michael Guerrette, George~S. Nolas, and
  Timothy~A. Strobel.
\newblock {BC8} silicon ({Si-III}) is a narrow-gap semiconductor.
\newblock {\em Physical Review Letter}, 118:146601, Apr 2017.

\bibitem{WongPRL2019}
S.~Wong, B.~Haberl, B.~C. Johnson, A.~Mujica, M.~Guthrie, J.~C. McCallum, J.~S.
  Williams, and J.~E. Bradby.
\newblock Formation of an r8-dominant si material.
\newblock {\em Phys. Rev. Lett.}, 122:105701, Mar 2019.

\bibitem{HaugeNanolett2015}
Håkon Ikaros~T. Hauge, Marcel~A. Verheijen, Sonia Conesa-Boj, Tanja
  Etzelstorfer, Marc Watzinger, Dominik Kriegner, Ilaria Zardo, Claudia
  Fasolato, Francesco Capitani, Paolo Postorino, Sebastian Kölling, Ang Li,
  Simone Assali, Julian Stangl, and Erik P. A.~M. Bakkers.
\newblock Hexagonal {Silicon} {Realized}.
\newblock {\em Nano Letters}, 15(9):5855--5860, September 2015.
\newblock Publisher: American Chemical Society.

\bibitem{PandolfiNanoLetters2018}
Silvia Pandolfi, Carlos Renero-Lecuna, Yann Le~Godec, Benoit Baptiste, Nicolas
  Menguy, Michele Lazzeri, Christel Gervais, Kristina Spektor, Wilson~A.
  Crichton, and Oleksandr~O. Kurakevych.
\newblock Nature of hexagonal silicon forming via high-pressure synthesis:
  Nanostructured hexagonal 4{H} polytype.
\newblock {\em Nano Letters}, 18(9):5989--5995, 2018.

\bibitem{BarthChemMat2020}
Sven Barth, Michael~S. Seifner, and Stephen Maldonado.
\newblock Metastable group iv allotropes and solid solutions: Nanoparticles and
  nanowires.
\newblock {\em Chemistry of Materials}, 32(7):2703--2741, 2020.

\bibitem{MalonePRB2008}
Brad~D. Malone, Jay~D. Sau, and Marvin~L. Cohen.
\newblock Ab initio survey of the electronic structure of tetrahedrally bonded
  phases of silicon.
\newblock {\em Phys. Rev. B}, 78:035210, Jul 2008.

\bibitem{BradbyJMR2001}
J.~E. Bradby, J.~S. Williams, J.~Wong-Leung, M.~V. Swain, and P.~Munroe.
\newblock Mechanical deformation in silicon by micro-indentation.
\newblock {\em Journal of Materials Research}, 16(5):1500–1507, 2001.

\bibitem{KailerJAP1997}
A.~Kailer, Y.~G. Gogotsi, and K.~G. Nickel.
\newblock Phase transformations of silicon caused by contact loading.
\newblock {\em Journal of Applied Physics}, 81(7):3057--3063, 1997.

\bibitem{JangActaMat2005}
Jae-il Jang, M.J. Lance, Songqing Wen, Ting~Y. Tsui, and G.M. Pharr.
\newblock Indentation-induced phase transformations in silicon: influences of
  load, rate and indenter angle on the transformation behavior.
\newblock {\em Acta Materialia}, 53(6):1759--1770, 2005.

\bibitem{HustonNanoLett2021}
Larissa~Q. Huston, Alois Lugstein, Guoyin Shen, David~A. Cullen, Bianca Haberl,
  Jim~S. Williams, and Jodie~E. Bradby.
\newblock Synthesis of {Novel} {Phases} in {Si} {Nanowires} {Using} {Diamond}
  {Anvil} {Cells} at {High} {Pressures} and {Temperatures}.
\newblock {\em Nano Letters}, 21(3):1427--1433, February 2021.
\newblock Publisher: American Chemical Society.

\bibitem{PandolfiNatComm2022}
Silvia Pandolfi, S.~Brennan Brown, P.~G. Stubley, Andrew Higginbotham, C.~A.
  Bolme, H.~J. Lee, B.~Nagler, E.~Galtier, R.~L. Sandberg, W.~Yang, W.~L. Mao,
  J.~S. Wark, and A.~E. Gleason.
\newblock Atomistic deformation mechanism of silicon under laser-driven shock
  compression.
\newblock {\em Nature Communications}, 13(1):5535, September 2022.

\bibitem{KiranBOOK2015}
Mangalampalli~S.R.N. Kiran, Bianca Haberl, Jodie~E. Bradby, and James~S.
  Williams.
\newblock Chapter five - nanoindentation of silicon and germanium.
\newblock In Lucia Romano, Vittorio Privitera, and Chennupati Jagadish,
  editors, {\em Defects in Semiconductors}, volume~91 of {\em Semiconductors
  and Semimetals}, pages 165--203. Elsevier, 2015.

\bibitem{GerbigJMR2015}
Yvonne~B. Gerbig, Chris~A. Michaels, and Robert~F. Cook.
\newblock In situ observation of the spatial distribution of crystalline phases
  during pressure-induced transformations of indented silicon thin films.
\newblock {\em Journal of Materials Research}, 30(3):390--406, February 2015.

\bibitem{DomnichRevAdvMatSci2002}
V.~Domnich and Y.~Gogotsi.
\newblock Phase transformations in silicon under contact loading.
\newblock {\em Reviews of Advanced Materials Science}, 3:1--36, 2002.

\bibitem{LiangScriptaMat2022}
Tao Liang, Lianghua Xiong, Hongbo Lou, Fujun Lan, Junran Zhang, Ye~Liu,
  Dongsheng Li, Qiaoshi Zeng, and Zhidan Zeng.
\newblock Mechanical properties of hexagonal silicon.
\newblock {\em Scripta Materialia}, 220:114936, November 2022.

\bibitem{WongJAP2019}
S.~Wong, B.~C. Johnson, B.~Haberl, A.~Mujica, J.~C. McCallum, J.~S. Williams,
  and J.~E. Bradby.
\newblock Thermal evolution of the indentation-induced phases of silicon.
\newblock {\em Journal of Applied Physics}, 126(10):105901, September 2019.

\bibitem{bartok2018machine}
Albert~P Bart{\'o}k, James Kermode, Noam Bernstein, and G{\'a}bor Cs{\'a}nyi.
\newblock Machine learning a general-purpose interatomic potential for silicon.
\newblock {\em Physical Review X}, 8(4):041048, 2018.

\bibitem{LysogorskiyNPJCompMat2021}
Yury Lysogorskiy, Cas van~der Oord, Anton Bochkarev, Sarath Menon, Matteo
  Rinaldi, Thomas Hammerschmidt, Matous Mrovec, Aidan Thompson, Gábor Csányi,
  Christoph Ortner, and Ralf Drautz.
\newblock Performant implementation of the atomic cluster expansion ({PACE})
  and application to copper and silicon.
\newblock {\em npj Computational Materials}, 7(1):97, June 2021.

\bibitem{GeActaMat2024}
Guojia Ge, Fabrizio Rovaris, Daniele Lanzoni, Luca Barbisan, Xiaobin Tang, Leo
  Miglio, Anna Marzegalli, Emilio Scalise, and Francesco Montalenti.
\newblock Silicon phase transitions in nanoindentation: Advanced molecular
  dynamics simulations with machine learning phase recognition.
\newblock {\em Acta Materialia}, 263:119465, 2024.

\bibitem{SheppardJCP2012}
Daniel Sheppard, Penghao Xiao, William Chemelewski, Duane~D. Johnson, and
  Graeme Henkelman.
\newblock A generalized solid-state nudged elastic band method.
\newblock {\em Journal of Chemical Physics}, 136(7), 2012.

\bibitem{HenkelmanJCP2000}
Graeme Henkelman and Hannes Jónsson.
\newblock {Improved tangent estimate in the nudged elastic band method for
  finding minimum energy paths and saddle points}.
\newblock {\em The Journal of Chemical Physics}, 113(22):9978--9985, 12 2000.

\bibitem{ThompsonCPC2022}
A.~P. Thompson, H.~M. Aktulga, R.~Berger, D.~S. Bolintineanu, W.~M. Brown,
  P.~S. Crozier, P.~J. in~'t Veld, A.~Kohlmeyer, S.~G. Moore, T.~D. Nguyen,
  R.~Shan, M.~J. Stevens, J.~Tranchida, C.~Trott, and S.~J. Plimpton.
\newblock {LAMMPS} - a flexible simulation tool for particle-based materials
  modeling at the atomic, meso, and continuum scales.
\newblock {\em Comp. Phys. Comm.}, 271:108171, 2022.

\bibitem{stukowski2009visualization}
Alexander Stukowski.
\newblock Visualization and analysis of atomistic simulation data with
  ovito--the open visualization tool.
\newblock {\em Modelling and Simulation in Materials Science and Engineering},
  18(1):015012, 2009.

\bibitem{ParrinelloJAP1981}
M.~Parrinello and A.~Rahman.
\newblock Polymorphic transitions in single crystals: {A} new molecular
  dynamics method.
\newblock {\em Journal of Applied Physics}, 52(12):7182--7190, December 1981.

\bibitem{TSASE}
http://theory.cm.utexas.edu/henkelman/code/.
\newblock To obtain the TSASE code.

\bibitem{HenkelmanJCP2000_2}
Graeme Henkelman, Blas~P. Uberuaga, and Hannes Jónsson.
\newblock {A climbing image nudged elastic band method for finding saddle
  points and minimum energy paths}.
\newblock {\em The Journal of Chemical Physics}, 113(22):9901--9904, 12 2000.

\bibitem{TherrienPRL2020}
F\'elix Therrien and Vladan Stevanovi\ifmmode~\acute{c}\else \'{c}\fi{}.
\newblock Minimization of atomic displacements as a guiding principle of the
  martensitic phase transformation.
\newblock {\em Phys. Rev. Lett.}, 125:125502, Sep 2020.

\bibitem{VineyardJPCS1957}
George~H. Vineyard.
\newblock Frequency factors and isotope effects in solid state rate processes.
\newblock {\em Journal of Physics and Chemistry of Solids}, 3(1-2):121--127,
  January 1957.

\bibitem{YinnpjCompMat2020}
Ketao Yin, Pengyue Gao, Xuecheng Shao, Bo~Gao, Hanyu Liu, Jian Lv, John~S. Tse,
  Yanchao Wang, and Yanming Ma.
\newblock An automated predictor for identifying transition states in solids.
\newblock {\em npj Computational Materials}, 6(1):16, March 2020.

\bibitem{WangPRL2013}
Jian-Tao Wang, Changfeng Chen, Hiroshi Mizuseki, and Yoshiyuki Kawazoe.
\newblock Kinetic {Origin} of {Divergent} {Decompression} {Pathways} in
  {Silicon} and {Germanium}.
\newblock {\em Physical Review Letters}, 110(16):165503, April 2013.

\bibitem{JiapengSciRep2017}
Sun Jiapeng, Li~Cheng, Jing Han, Aibin Ma, and Liang Fang.
\newblock Nanoindentation {Induced} {Deformation} and {Pop}-in {Events} in a
  {Silicon} {Crystal}: {Molecular} {Dynamics} {Simulation} and {Experiment}.
\newblock {\em Scientific Reports}, 7(1), December 2017.
\newblock Publisher: Nature Publishing Group.

\bibitem{abram2022silicon}
Rafa{\l} Abram, Dariusz Chrobak, Jesper Byggm{\"a}star, Kai~H Nordlund, and
  Roman Nowak.
\newblock Silicon nanoindentation modelled by hybrid potential.
\newblock {\em arXiv preprint arXiv:2207.08102}, 2022.

\bibitem{BellScience1978}
George~I. Bell.
\newblock Models for the specific adhesion of cells to cells.
\newblock {\em Science}, 200(4342):618 – 627, 1978.

\bibitem{GhasemiJMPS2020}
Arman Ghasemi and Wei Gao.
\newblock A method to predict energy barriers in stress modulated solid–solid
  phase transitions.
\newblock {\em Journal of the Mechanics and Physics of Solids}, 137:103857,
  April 2020.

\bibitem{LevitasPRB2017}
Valery~I. Levitas, Hao Chen, and Liming Xiong.
\newblock Lattice instability during phase transformations under multiaxial
  stress: {Modified} transformation work criterion.
\newblock {\em Physical Review B}, 96(5):054118, August 2017.

\bibitem{LevitasPRL2017}
Valery~I. Levitas, Hao Chen, and Liming Xiong.
\newblock Triaxial-{Stress}-{Induced} {Homogeneous} {Hysteresis}-{Free}
  {First}-{Order} {Phase} {Transformations} with {Stable} {Intermediate}
  {Phases}.
\newblock {\em Physical Review Letters}, 118(2), January 2017.
\newblock Publisher: American Physical Society.

\bibitem{SchaffarJOM2022}
Gerald J.~K. Schaffar, Johann Kappacher, Daniel Tscharnuter, and Verena
  Maier-Kiener.
\newblock The {Phase} {Transformation} of {Silicon} {Assessed} by an
  {Unloading} {Contact} {Pressure} {Approach}.
\newblock {\em JOM}, 74(6):2220--2230, June 2022.

\bibitem{LinPRL2020}
Chuanlong Lin, Xuqiang Liu, Dongliang Yang, Xiaodong Li, Jesse~S. Smith, Bihan
  Wang, Haini Dong, Shourui Li, Wenge Yang, and John~S. Tse.
\newblock Temperature- {A} nd {Rate}-{Dependent} {Pathways} in {Formation} of
  {Metastable} {Silicon} {Phases} under {Rapid} {Decompression}.
\newblock {\em Physical Review Letters}, 125(15), October 2020.

\bibitem{MannepalliJAP2019}
Sowjanya Mannepalli and Kiran S. R.~N. Mangalampalli.
\newblock {In-situ high temperature micro-Raman investigation of annealing
  behavior of high-pressure phases of Si}.
\newblock {\em Journal of Applied Physics}, 125(22):225105, 06 2019.

\bibitem{PiltzPRB1995}
R.~O. Piltz, J.~R. Maclean, S.~J. Clark, G.~J. Ackland, P.~D. Hatton, and
  J.~Crain.
\newblock Structure and properties of silicon {XII}: {A} complex tetrahedrally
  bonded phase.
\newblock {\em Physical Review B}, 52(6):4072--4085, August 1995.

\bibitem{RuffellAPL2007}
S.~Ruffell, J.~E. Bradby, and J.~S. Williams.
\newblock Annealing kinetics of nanoindentation-induced polycrystalline high
  pressure phases in crystalline silicon.
\newblock {\em Applied Physics Letters}, 90(13):131901, March 2007.

\bibitem{RuffellJAP2009}
S.~Ruffell, B.~Haberl, S.~Koenig, J.~E. Bradby, and J.~S. Williams.
\newblock Annealing of nanoindentation-induced high pressure crystalline phases
  created in crystalline and amorphous silicon.
\newblock {\em Journal of Applied Physics}, 105(9):093513, May 2009.

\bibitem{GeJAP2004}
Daibin Ge, Vladislav Domnich, and Yury Gogotsi.
\newblock Thermal stability of metastable silicon phases produced by
  nanoindentation.
\newblock {\em Journal of Applied Physics}, 95(5):2725--2731, March 2004.

\bibitem{BlowJCP2021}
Katarina~E. Blow, David Quigley, and Gabriele~C. Sosso.
\newblock The seven deadly sins: {When} computing crystal nucleation rates, the
  devil is in the details.
\newblock {\em The Journal of Chemical Physics}, 155(4):040901, July 2021.

\end{thebibliography}

\end{document}